\documentclass[submission,copyright,creativecommons]{eptcs}
\providecommand{\event}{Infinity 2012} 
\usepackage{times}
\usepackage{psfig}
\usepackage{graphics}
\usepackage{graphicx}
\usepackage{xspace}
\usepackage{amsmath,amssymb}
\usepackage{amsfonts}
\usepackage{algorithm}
\usepackage{algorithmic}
\usepackage{array}
\usepackage{gastex}

\usepackage{psfrag} 
\usepackage{subfigure}
\usepackage{stfloats}
\usepackage{cite}
\usepackage{color}
\usepackage{paralist}
\usepackage{boxedminipage}
\usepackage{mdwmath}
\usepackage{mdwtab}

\newcommand{\Nat}{{\mathbb N}}
\usepackage[all,2cell,ps]{xy}

\title{On Consistency of Operational Transformation Approach}
\author{Aurel~Randolph$^1$,
        Hanifa~Boucheneb$^1$,
        Abdessamad~Imine$^2$,
        and~Alejandro~Quintero$^1$ \\ 
 $^1$\'{E}cole Polytechnique de Montr\'{e}al, \\ P.O. Box 6079, Station Centre-ville, Montr\'{e}al, Qu\'{e}bec, Canada, H3C 3A7. \email{ \{aurel.randolph, hanifa.boucheneb, alejandro.quintero\}@polymtl.ca} \\ 
 $^2$INRIA Grand-Est and Nancy-Universit\'e, France. \email{imine@loria.fr}
}

\def\titlerunning{On Consistency of Operational Transformation Approach}
\def\authorrunning{A. Randolph, H. Boucheneb, A. Imine, A. Quintero}
\begin{document}
\maketitle

\begin{abstract} 
The Operational Transformation (OT) approach, used in many collaborative editors, allows a group of 
users to concurrently update replicas of a shared object and exchange their updates in any order. 
The basic idea of this approach is to transform any received update operation before its execution 
on a replica of the object. This transformation aims to ensure the convergence of the different replicas 
of the object, even though the operations are executed in different orders. However, designing transformation 
functions for achieving convergence is a critical and challenging issue. Indeed, the transformation 
functions proposed in the literature are all revealed incorrect. \\
\par In this paper, we investigate 
the existence of transformation functions for a shared string altered by insert and delete operations. 
From the theoretical point of view, two properties -- named TP1 and TP2 -- are necessary and sufficient 
to ensure convergence. 
Using controller synthesis technique, we show that there are some transformation functions
which satisfy only TP1 for the basic signatures of insert and delete operations.
As a matter of fact, it is impossible to meet both properties TP1 and TP2 with these simple signatures.


\end{abstract}

\section{Introduction} \label{intro}
\par Collaborative editing systems (CESs for short) constitute a class of distributed systems where dispersed users interact by manipulating some shared objects like texts, images, graphics, XML documents, etc. To improve data availability, these systems are based on data replication. Each user has its local copy of the shared object and can access and update its local copy. The update operations executed locally are propagated to other users. Update operations are not necessarily executed in the same order on the object replicas, which may lead to a divergence (object replicas are not identical). For instance, suppose two users $u_1$ and $u_2$ working on their own copies of a text containing the word \emph{``efecte''}.  User $u_1$ inserts $`f'$ at position $1$, to change the word into \emph{``effecte''}. Concurrently, user $u_2$ deletes element at position $5$ (i.e., the last $'e'$), to change the word into \emph{``efect''}.  Each user will receive an update operation that was applied on a different version of the text. Applying naively the received update operations will lead to divergent replicas (\emph{``effece''} for user $u_1$ and \emph{``effect''} for user $u_2$, see Fig.\ref{fig:incons}).\\

\par Several approaches are proposed in the literature, to deal with the convergence of replicated data: Multi-Version (MV), Serialization-Resolution of Conflicts (SRC), Commutative Replicated  Data Type (CRDT),  Operational Transformation (OT), etc. \\ 

\par The multi-version approach  \cite{Ber83}, used in CVS, Subversion and ClearCase, is based on the paradigm ``Copy-Modify-Merge''. In this approach, update operations made by a user are not automatically propagated to the others. They will be propagated only when the user call explicitly the merge function. It would be interesting to propagate automatically, to all others, each update operation performed by a user. This is the basic idea of SRC.\\ 

\par To achieve convergence, SRC imposes to execute the operations in the same order at every site. Therefore, sites may have to undo and execute again operations, as they receive the final execution order of update operations. This order is determined by a central server fixed when the system is launched (central node). For the previous example, this approach requires that sites of both users execute the two operations in the same order. However, even if we obtain an identical result in both sites, the execution order imposed by the central site may not correspond to the original intention of some user. For instance, executing, in both sites,  the operation of $u_1$ followed by the one of $u_2$ results in the text \emph{``effece''}, which  is  inconsistent with the intention of $u_2$.\\

\par The Commutative Replicated Data Type (CRDT) is a data type where all concurrent operations commute with each other \cite{PRE09}.  In such a case, to ensure convergence of replicas it suffices to respect the causality principle (i.e., whenever an operation $o'$ is generated after executing another operation $o$, $o$ is executed before $o'$ at every site). The main challenge of CRDT is designing  commutative operations for the data type.  The commonly used idea consists in associating a unique identifier with the position of each symbol, line or atom of the shared  document and when an insert operation is generated, a unique identifier is also associated with the position parameter of the operation.  The position identifiers  do not change and are totally ordered w.r.t. $<$. Symbols, lines or atoms of the document appear in increasing order w.r.t. their identifiers.  Managing position identifiers is a very important issue in this approach as the correctness is based on the unicity of position identifiers and the total order preservation.  Ensuring unicity may induce  space and time overheads.  Let us apply this paradigm to the previous example. A unique identifier is associated with each symbol of the initial text: \emph{``(e,3) (f,6) (e, 8) (c,9) (t,9.5) (e,10)''}. A unique identifier  between $3$ and $6$ is affected to position $1$ of the operation of $u_1$. Let $4.5$ be the selected identifier. The identifier affected to position $5$ of the delete operation of $u_2$ is $10$.  Both execution orders of operations of $u_1$ and $u_2$ lead to the text \emph{``(e,3) (f,4.5) (f,6) (e, 8) (c,9) (t,9.5)''}.  CESs like TreeDoc \cite{PRE09}, Logoot \cite{wei09},  Logoot-Undo \cite{Wei10}  and  WOOT \cite{Ost06} are based on CRDT paradigm.  In this approach, all concurrent operations are commutative. So, the different orders of their execution lead to the same state. \\

\par Operational transformation (OT) proposed by \cite{Ellis89} is an approach where the generated concurrent operations are not necessarily commutative. Their commutativity is forced by transformation of operations before their execution. More precisely, when a site receives an update operation, it is first transformed w.r.t. concurrent operations already executed on the site. The transformed operation is then executed on the local copy. This transformation aims at assuring the convergence of copies even if users execute the same set of operations in different orders. OT is based on a transformation function, called Inclusive Transformation (IT), which transforms an update operation w.r.t. another update operation.  For the  previous example, when $u_1$ receives the operation of $u_2$, it is first transformed  w.r.t. the local operation as follows: $IT(Del(5),Ins(1,\mbox{\emph{f}}))  = Del(6)$. The deletion position  is incremented  because $u_1$ has inserted a character at position $1$, which  is before the character deleted by  $u_2$. Next, the transformed operation is  executed on the local copy
 of $u_1$. In a similar way, when $u_2$ receives the operation of $u_1$, it is transformed as follows before its execution on the local copy of $u_2$:  $IT(Ins(1,\mbox{\emph{f}}),Del(5))= Ins(1,\mbox{\emph{f}})$. In this case, it remains  the  same  because \emph{f} is inserted before the deletion position of operation of $u_2$ (see Fig.\ref{fig:transf}). We can find, in the literature, several IT functions: \emph{Ellis}'s algorithm \cite{Ellis89}, \emph{Ressel}'s algorithm \cite{Ressel.ea:96}, \emph{Sun}'s algorithm \cite{Sun.ea:98}, \emph{Suleiman}'s algorithm \cite{suleiman97} and  \emph{Imine}'s algorithm \cite{imine02b}. However, all these functions fail to ensure convergence \cite{Imi06,BI09, IFM10}.  \\ 
\par In this paper,  we investigate the existence of IT functions ensuring convergence for shared strings based on the classical signatures of update operations.  Section \ref{s2} is devoted to OT and IT functions proposed in the literature. For each IT function, we provide, at this level, a counterexample for the convergence property. In Section \ref{s3}, we show, using a controller synthesis technique, that there is no IT function based on the classical signatures of update operations, which ensures convergence. Conclusion goes in Section \ref{s4}.

\begin{figure}[htbp]
\centering
\scriptsize
 \begin{minipage}[t]{0.9\linewidth}
\centerline{\xymatrix@C=10pt@M=2pt@R=10pt{ *+[F-,]\txt{site 1 \\
``efecte''} \ar@{.}'[d]'[dd]'[ddd][dddd] &
*+[F-,]\txt{site 2 \\ ``efecte''} \ar@{.}'[d]'[dd]'[ddd][dddd] \\
o_1=Ins(1,f) \ar[ddr]  & o_2=Del(5) \ar[ddl] |!{[l];[dd]}\hole \\
*+[F]{\txt{``effecte''}} & *+[F]{\txt{``efect''}} \\
  Del(5)     &   Ins(1,f) \\
*+[F]{\txt{``effece''}}  & *+[F]{\txt{``effect''}} \\
}} 
  \caption{Integration without transformation.}
  \label{fig:incons}
\hfill
 \end{minipage}

 \begin{minipage}[t]{0.9\linewidth}
\centerline{\xymatrix@C=20pt@M=2pt@R=10pt{ *+[F-,]\txt{site 1 \\
``efecte''} \ar@{.}'[d]'[dd]'[ddd][dddd] &
*+[F-,]\txt{site 2 \\ ``efecte''} \ar@{.}'[d]'[dd]'[ddd][dddd] \\
o_1=Ins(1,f) \ar[ddr]  & o_2=Del(5) \ar[ddl] |!{[l];[dd]}\hole \\
*+[F]{\txt{``effecte''}} & *+[F]{\txt{``efect''}} \\
IT(o_2,o_1)=Del(6)     & IT(o_1,o_2)=Ins(1,f) \\
*+[F]{\txt{``effect''}}  & *+[F]{\txt{``effect''}} \\}}
  \caption{Integration with transformation.}
  \label{fig:transf}
 \end{minipage} \normalsize
\end{figure}

\section{Operational Transformation Approach}\label{s2} 
\subsection{Background}\label{Def:caus}
 OT considers $n$ sites,  where  each  site  has   a  copy  of  the
collaborative  object (shared object). The shared object is a finite sequence of elements from a data type $\mathcal{A}$ (alphabet). 
It is assumed here that the shared object can only be modified by the following primitive operations:\\
\centerline{$\mathcal{O} =\{Ins(p,c) | c\in \mathcal{A}\mbox{ and } p\in\Nat\}
\cup \{Del(p) | p\in\Nat\} \cup \{Nop()\}$}\\
where $Ins(p,c)$ inserts the element $c$ at position $p$; $Del(p)$  deletes the element at position $p$, and $Nop()$ is the idle operation that has null effect on the object.\\ 
\par Each site can concurrently update its copy of the shared object. Its local updates are then propagated to other sites. When a site receives an update operation, it is first transformed before its execution. Since  the shared object is replicated, each site will own
a local state $l$ that is altered only by operations executed locally.
The initial state of the shared object, denoted $l_0$, is the same for all sites.
Let $\mathcal{L}$ be the set of states. The function $Do: \mathcal{O}\times\mathcal{L}\rightarrow
\mathcal{L}$, computes the state $Do(o,l)$ resulting from applying
operation $o$ to state $l$. We denote by $[o_1;o_2;\ldots;o_m]$ an operation sequence. Applying  an
operation sequence to a state $l$ is defined as follows:
\begin{inparaenum}[(i)]
\item $Do([],l)=l$, where $[]$ is the empty sequence and; \item
$Do([S;o],l)=Do(o,Do(S,l))$, $S$ being an operation sequence.\\
\end{inparaenum}
\par Two operation  sequences $S_1$ and  $S_2$ are
\textit{equivalent}, denoted $S_1\equiv S_2$, iff
$Do(S_1,l) =  Do(S_2,l)$ for all states $l$.\\

\par Concretely, OT consists of the integration procedure and the transformation function, called Inclusive Transformation (IT function). The integration procedure is in charge of executing update operations, broadcasting local update operations to other sites, receiving update operations from other sites, and determining transformations to be performed on a received operation before its execution. The transformation function transforms an update operation $o$ w.r.t. another update operation $o'$ ($IT(o,o')$). Let $S=[o_1;o_2;\ldots;o_m]$  be  a sequence of operations. Transforming any editing operation  $o$ w.r.t. $S$ is  denoted $IT^*(o,S)$ and is recursively defined by: 
$IT^*(o,[])=o\text{, where } [] \text{ is the empty sequence, and } IT^*(o,[o_1;o_2;\ldots;o_m])=IT^*(IT(o,o_1),[o_2;\ldots;o_m]).$ By definition: $IT(Nop(),o)=Nop()$ and $IT(o,Nop())=o$ for every operation  $o$. 

\subsection{Integration procedures}\label{sec:integration} 
 The integration procedure is based on two notions: concurrency and dependency of operations. Let $o_1$ and $o_2$ be two operations generated at sites $i$ and $j$, respectively. We say that $o_2$ \emph{causally
depends} on $o_1$, denoted $o_1 \rightarrow o_2$, iff:
\begin{inparaenum}[(i)]
\item $i=j$ and $o_1$ was generated before $o_2$; or, \item $i\neq
j$ and the execution of $o_1$ at site $j$ has happened before
      the generation of $o_2$.
\end{inparaenum} Operations $o_1$ and $o_2$ are said to be \emph{concurrent},
 denoted $o_1 \parallel o_2$, iff neither $o_1 \rightarrow o_2$ nor $o_2 \rightarrow o_1$.
\noindent As a long established convention in OT-based
collaborative editors~\cite{Ellis89,Sun98}, the \emph{timestamp
vectors} are used to determine the causality and concurrency
relations between operations. A timestamp vector is associated with each site and each generated operation. Every timestamp is a vector of integers with a number of entries equal to the number of sites. For a site $j$, each entry $V_j[i]$ returns the number of operations generated at site $i$ that have been already executed on site $j$. When an operation $o$ is generated at site $i$, a copy $V_o$ of $V_i$ is associated with $o$ before its broadcast to other sites. The entry $V_i[i]$ is then incremented by $1$. Once $o$ is received at site $j$, if the local vector $V_{j}$ ``dominates''\footnote{We say that $V_1$ dominates $V_2$ iff $\forall$ $i$, $V_1[i]\geq V_2[i]$.} $V_o$, then $o$ is ready to be executed on site $j$. In this case, $V_{j}[i]$ will be incremented by $1$ after the execution of $o$. Otherwise, the $o$'s execution is delayed. Let $V_{o_1}$ and $V_{o_2}$ be timestamp vectors of $o_1$ and $o_2$, respectively. Using these timestamp vectors, the causality and concurrency relations are defined as follows:
\begin{inparaenum}[(i)]
\item $o_1 \rightarrow o_2$ iff $V_{o_1}[i] <
V_{o_2}[j]$; \item $o_1 \parallel o_2$ iff $V_{o_1}[i]
\geq V_{o_2}[j]$
  and $V_{o_2}[i] \geq V_{o_1}[j]$.\\
\end{inparaenum}

\par Several integration procedures have been proposed in the
groupware research area, such as dOPT~\cite{Ellis89},
adOPTed~\cite{Ressel.ea:96}, SOCT2,4 \cite{Suleiman.ea:98,Vidot.ea:00}, GOTO~\cite{Sun98} and COT~\cite{Sun09}. There are two kinds of integration procedures: centralized and decentralized. In the centralized integration procedures such as SOCT4 and COT, there is a central node which ensures that all concurrent operations are executed in the same order at all sites. 
 In the decentralized integration procedures such as adOPTed, SOCT2 and GOTO,  there is no central node and the operations may be executed in different orders by different sites.   We focus, in the following, on the decentralized integration procedures.  In general, in such a kind of integration procedures, every site generates operations sequentially and stores these operations
in a stack also called a \textit{history} (or \emph{execution
trace}). When a site receives a remote operation $o$,  the
integration  procedure executes  the  following steps:
\begin{enumerate} 
\item From the local history $S$, it determines the equivalent
      sequence $S'$ that is the concatenation of two sequences
      $S_h$ and $S_c$ where (i) $S_h$ contains all operations
      happened before $o$ (according to the causality relation defined above), and (ii)
      $S_c$ consists of operations that are concurrent to $o$.
          \item It calls the transformation component in order to get
    operation $o'$ that is the transformation of $o$ according to
    $S_c$ (\textit{i.e.} $o'=IT^*(o,S_c)$).
    \item It executes $o'$ on the current state and then adds $o'$ to local history $S$.
\end{enumerate} 
The integration procedure allows history of executed
operations to be built on every site, provided that the causality
relation is preserved. When all sites have executed the same set of operations (stable states), their histories are not
necessarily identical because the concurrent operations may be
executed in different orders. Nevertheless, they must
be equivalent in the sense that they must lead to the same final
state.  
\subsection{Inclusive transformation functions}
We can find, in the literature, several IT functions: \emph{Ellis}'s algorithm \cite{Ellis89}, \emph{Ressel}'s algorithm
\cite{Ressel.ea:96}, \emph{Sun}'s algorithm \cite{Sun.ea:98}, \emph{Suleiman}'s algorithm \cite{suleiman97} and  \emph{Imine}'s
algorithm \cite{imine02b}. They differ in the manner that conflict situations are managed. A conflict situation occurs when two concurrent operations insert different characters at the same position. To deal with such conflicts, all these algorithms, except the one proposed by Sun et al., add some extra parameters to the insert operation signature.
\subsubsection{Ellis's algorithm} 
Ellis and Gibbs~\cite{Ellis89} are the pioneers of OT approach.
They extend  operation  $Ins$ with  another  parameter   $pr$  representing its priority. Concurrent operations have always different priorities.  Fig.\ref{alg:Ellis} illustrates the four transformation
cases  for $Ins$  and  $Del$  proposed by Ellis and Gibbs.
\begin{figure}[htbp]
\centering
\begin{boxedminipage}{0.9\linewidth}
\scriptsize
IT($Ins(p_1,c_1,pr_1),Ins(p_2,c_2,pr_2)$) = \\
$ \begin{cases} Ins(p_1,c_1,pr_1) & \rm{if} \ {(p_1<p_2) \vee }\\
&{(p_1=p_2 \wedge c_1 \neq c_2 \wedge pr_1 < pr_2) }
\\ Ins(p_1+1,c_1,pr_1) & \rm{if} \ { p_1>p_2 \vee }\\
&{(p_1=p_2 \wedge c_1 \neq c_2) \wedge pr_1> pr_2) } \\
Nop()  & \rm{otherwise}\\
\end{cases} $\\
IT($Ins(p_1,c_1,pr_1), Del(p_2)$)= $ \begin{cases}
Ins(p_1,c_1,pr_1) & \rm{if}
 \ {p_1 < p_2} \\
Ins(p_1-1,c_1,pr_1) & \rm{otherwise}\\
\end{cases} $\\
IT($Del(p_1),Ins(p_2,c_2,pr_2)$) =$ \begin{cases} Del(p_1) & \rm{if} \  {p_1<p_2} \\
Del(p_1+1) & \rm{otherwise}\\
\end{cases} $\\
IT($Del(p_1),Del(p_2)$) = $ \begin{cases} Del(p_1) & \rm{if} \  {p_1<p_2} \\
Del(p_1-1) & \rm{if} \  {p_1>p_2} \\
Nop() & \rm{otherwise} \\
\end{cases} $ \normalsize
\end{boxedminipage}
\caption{IT function of Ellis et \textit{al}.}
\label{alg:Ellis}
\end{figure}
\subsubsection{Ressel's algorithm}
Ressel et \textit{al.}~\cite{Ressel.ea:96} proposed an algorithm that
provides two modifications in Ellis's algorithm. The first modification consists in
replacing priority parameter $pr$ by another parameter $u$, which is
simply the \emph{identifier} of the issuer site. Similarly, $u$ is
used for tie-breaking when a conflict occurs between two concurrent
insert operations. As for the second modification, it concerns how a pair of  insert
operations is transformed. When two concurrent insert operations add
at the same position two (identical or different) elements, only the
insertion position of operation having a higher identifier is
incremented. In other words, the both elements are inserted even if
they are identical. What is opposite to solution proposed by Ellis and
Gibbs, which keeps only one element in case of identical concurrent
insertions. Apart from these modifications, the other cases remain
similar to those of Ellis and Gibb. Fig.~\ref{alg:Ressel} illustrates all transformation cases given
by the algorithm of Ressel et \textit{al.}~\cite{Ressel.ea:96}.
\begin{figure}[htbp]
\centering
\begin{boxedminipage}{0.9\linewidth}
\scriptsize
IT($Ins(p_1,c_1,u_1),Ins(p_2,c_2,u_2)$) =
$ \begin{cases} Ins(p_1,c_1,u_1) & \rm{if} \ {p_1<p_2 \vee
(p_1=p_2  \wedge u_1 < u_2) }
\\ Ins(p_1+1,c_1,u_1) & \rm{otherwise} \\
\end{cases} $\\
IT($Ins(p_1,c_1,u_1), Del(p_2)$)= $ \begin{cases}
Ins(p_1,c_1,u_1) & \rm{if}
 \ {p_1 \leq p_2} \\
Ins(p_1-1,c_1,u_1) & \rm{otherwise}\\
\end{cases} $\\
IT($Del(p_1),Ins(p_2,c_2,u_2)$) =
$ \begin{cases} Del(p_1) & \rm{if} \  p_1<p_2 \\
Del(p_1+1) & \rm{otherwise}\\
\end{cases} $\\
IT($Del(p_1),Del(p_2)$) = $ \begin{cases} Del(p_1) & \rm{if} \  {p_1<p_2} \\
Del(p_1-1) & \rm{if} \  {p_1>p_2} \\
Nop() & \rm{otherwise} \\
\end{cases} $ \normalsize
\end{boxedminipage}
\caption{IT function of Ressel et \textit{al}.}
\label{alg:Ressel}
\end{figure}
\subsubsection{Sun's algorithm} 
Sun et \textit{al}.~\cite{Sun.ea:98} have designed another IT
algorithm, which is slightly different in the sense that it is defined for stringwise
operations. Indeed, the following operations are used:  $Ins(p,s,l)$ to insert string $s$ of length $l$ at position $p$ and
 $Del(p,l)$ to delete string of length $l$ from position $p$. 
To compare with other IT algorithms, we suppose that $l=1$ for all update operations. The IT function in this case is reported at Fig.~\ref{alg:Sun}.
\begin{figure}[htbp]
\centering
\begin{boxedminipage}{0.9\linewidth}
\scriptsize
IT($Ins(p_1,c_1),Ins(p_2,c_2)$) =
$ \begin{cases} Ins(p_1,c_1) & \rm{if} \ {p_1<p_2}
\\ Ins(p_1+1,c_1) & \rm{otherwise}\\
\end{cases} $\\
IT($Ins(p_1,c_1), Del(p_2)$)= $ \begin{cases}
Ins(p_1,c_1) & \rm{if}
 \ {p_1 \leq p_2} \\
Ins(p_1-1,c_1) & \rm{otherwise}\\
\end{cases} $\\
IT($Del(p_1),Ins(p_2,c_2)$) =
$ \begin{cases} Del(p_1) & \rm{if} \  {p_1<p_2} \\
Del(p_1+1) & \rm{otherwise}\\
\end{cases} $\\
IT($Del(p_1),Del(p_2)$) = $ \begin{cases} Del(p_1) & \rm{if} \  {p_1<p_2} \\
Del(p_1-1) & \rm{if} \  {p_1>p_2} \\
Nop() & \rm{otherwise} \\
\end{cases} $ \normalsize
\end{boxedminipage}
\caption{Characterwise IT function of Sun et \textit{al}.}
\label{alg:Sun}
\end{figure}
\subsubsection{Suleiman's algorithm} 
Suleiman et \textit{al}.~\cite{suleiman97} proposed another solution
that modifies the signature of insert operation by adding two
parameters $av$ and $ap$. For an insert operation $Ins(p,c,av,ap)$,
$av$ contains operations that
have deleted a character before the insertion position $p$. The set
$ap$ contains operations that have removed a character after or at position $p$.
When an insert operation is generated the parameters $av$ and $ap$ are
empty. They will be filled during transformation steps. The $IT$ algorithms of Suleiman and \textit{al}. is given in
Figure~\ref{alg:Suleiman}. To resolve the conflict between two
concurrent insert operations $Ins(p,c_1,av_1,ap_1)$ and
$Ins(p,c_2,av_2,ap_2)$, three cases are possible: \\
1) \ $(av_1 \cap ap_2) \neq \emptyset$: character $c_2$ is inserted before
  character $c_1$, \\
2) \ $(ap_1 \cap av_2) \neq \emptyset$: character $c_2$ is inserted after
character $c_1$, \\
3) \ $(av_1 \cap ap_2) = (ap_1 \cap av_2) = \emptyset$: in this case
  characters $c_1$ and $c_2$ are compared (for instance according to the lexicographic order) to choose the one to be added before the other. Like the site identifiers and priorities, parameters $av$, $ap$, comparison of characters are used to tie-break
  conflict situations.
Note that when two concurrent operations insert the same character
(\textit{e.g.} $c_1=c_2$) at the same position, the one is
executed and the other one is ignored by returning the idle operation $Nop()$.
In other words, like the solution of Ellis and Gibb~\cite{Ellis89},
only one character is kept.
\begin{figure}[htbp]
\centering
\begin{boxedminipage}{0.95\linewidth}
\scriptsize
IT($Ins(p_1,c_1,av_1,ap_1),Ins(p_2,c_2,av_2,ap_2)$) = 
$ \begin{cases} Ins(p_1,c_1,,av_1,ap_1) &\rm{if} \ {p_1 < p_2}  \vee  \\ 
& (p_1 = p_2 \wedge ap_1 \cap av_2 \neq \emptyset) \vee \\
& (p_1 = p_2 \wedge ap_1 \cap av_2 = av_1 \cap ap_2 =\emptyset\\
&\wedge c_1 > c_2) \\ 

Ins(p_1+1,c_1,av_1,ap_1) & \rm{if} \ {p_1 > p_2}  \vee  \\ 
& (p_1 = p_2 \wedge av_1 \cap ap_2 \neq \emptyset) \vee \\ 
& (p_1 = p_2 \wedge ap_1 \cap av_2 = av_1 \cap ap_2 =\emptyset\\
&\wedge c_1 < c_2) \\ Nop() & \rm{otherwise}\\
\end{cases} $\\
IT($Ins(p_1,c_1,av_1,ap_1), Del(p_2)$)= $ \begin{cases}
Ins(p_1,c_1,av_1,ap_1\cup \{Del(p_2)\}) & \rm{if} \ {p_1\leq p_2} \\
Ins(p_1-1,c_1,av_1\cup \{Del(p_2)\},ap_1) & \rm{otherwise}\\
\end{cases} $\\
IT($Del(p_1),Ins(p_2,c_2,av_2,ap_2)$) =
$ \begin{cases} Del(p_1) & \rm{if} \  {p_1<p_2} \\
Del(p_1+1) & \rm{otherwise}\\
\end{cases} $\\
IT($Del(p_1),Del(p_2)$) = $ \begin{cases} Del(p_1) & \rm{if} \  {p_1<p_2} \\
Del(p_1-1) & \rm{if} \  {p_1>p_2} \\
Nop() & \rm{otherwise} \\
\end{cases} $ \normalsize
\end{boxedminipage}
\caption{IT function of Suleiman and \textit{al}.}
\label{alg:Suleiman} 
\end{figure}

\subsubsection{Imine's algorithm} 
In~\cite{imine02b}, Imine and \textit{al}. proposed another IT
algorithm which again enriches the signature of insert
operation with parameter $ip$ which is the initial (or the original)
insertion position given at the generation stage.
Thus, when transforming a pair of insert operations having the same
current position, they compare first their initial positions in order to
recover the position relation at the generation phase. If the initial
positions are identical, then like Suleiman and
\textit{al}.~\cite{suleiman97} they compare symbols to tie-break
an eventual conflict. Fig.~\ref{alg:Imine} gives
the $IT$ function of Imine.
\begin{figure}[htbp]
\centering
\begin{boxedminipage}{0.9\linewidth}
\footnotesize 
IT($Ins(p_1,c_1,ip_1),Ins(p_2,c_2,ip_2)$) = 
$ \begin{cases} Ins(p_1,c_1,ip_1) &\rm{if} \ {p_1 < p_2}  \vee  (p_1 = p_2 \wedge ip_1 < ip_2) \ \vee \\ & (p_1 = p_2 \wedge ip_1 = ip_2 \wedge c_1 < c_2) \\ 
 Ins(p_1+1,c_1,ip_1) & \rm{if} \ {p_1 > p_2}  \vee (p_1 = p_2 \wedge ip_1 > ip_2) \  \vee \\ & (p_1 = p_2 \wedge ip_1 = ip_2 \wedge c_1 > c_2) \\ Nop() & \rm{otherwise}\\
\end{cases} $\\
IT($Ins(p_1,c_1,ip_1), Del(p_2)$)= $ \begin{cases}
Ins(p_1,c_1,ip_1) & \rm{if} \ {p_1\leq p_2} \\
Ins(p_1-1,c_1,ip_1) & \rm{otherwise}\\
\end{cases} $\\
IT($Del(p_1),Ins(p_2,c_2,ip_2)$) =
$ \begin{cases} Del(p_1) & \rm{if} \  {p_1<p_2} \\
Del(p_1+1) & \rm{otherwise}\\
\end{cases} $\\
IT($Del(p_1),Del(p_2)$) = $ \begin{cases} Del(p_1) & \rm{if} \  {p_1<p_2} \\
Del(p_1-1) & \rm{if} \  {p_1>p_2} \\
Nop() & \rm{otherwise} \\
\end{cases} $ \normalsize
\end{boxedminipage}
\caption{IT function of Imine and \textit{al}.}
\label{alg:Imine} 
\end{figure}

\subsection{Consistency criteria} 

An OT-based collaborative editor is \emph{consistent} iff it satisfies the following properties:
\begin{enumerate}
\item \emph{Causality preservation:}
if $o_1 \rightarrow o_2$ then $o_1$ is
  executed before $o_2$ at all sites.
\item \emph{Convergence:} when all sites have performed the same
set of updates, the copies of the shared document are identical.
\end{enumerate} 
\indent To preserve the causal dependency between updates, timestamp vectors are used. 
In~\cite{Ressel.ea:96}, the authors have established two properties $TP1$  and $TP2$  that are necessary and sufficient to ensure data convergence for \textit{any number} of operations executed in \textit{arbitrary order} on copies of the same
object (i.e., decentralized integration procedure): For all $o_1$, $o_2$ and $o_3$ pairwise
concurrent operations generated on the same state (initial state or state reached from the initial state by executing equivalent sequences): 
\begin{itemize}[$\bullet$] 
\item \textbf{$TP1$}: $[o_1\,;IT(o_2,o_1)]\,\equiv\,
[o_2\,; IT(o_1,o_2)]$. \item \textbf{$TP2$}:
       $IT^*(o_3, [o_1\,;IT(o_2,o_1)])\,=\,IT^*(o_3, [o_2\,; IT(o_1,o_2)])$.
\end{itemize} 
Property $TP1$  defines a \emph{state identity} and ensures that
if $o_1$ and $o_2$ are concurrent, the effect of executing $o_1$
before $o_2$ is the  same  as  executing  $o_2$  before  $o_1$.
Property $TP2$ ensures that transforming $o_3$ along equivalent and different
operation sequences will give the same operation. By abuse of language, an IT function satisfying properties TP1 and TP2 is said be consistent.\\

\par Accordingly, by these properties, it is not necessary to enforce a global total order between concurrent operations because data
divergence can always be repaired by operational transformation.
However, finding an IT function that satisfies $TP1$ and $TP2$ is
considered as a hard task, because this proof is often
unmanageably complicated. Note that for some centralized integration procedures such as SOCT4 and COT, property TP1 is a necessary and sufficient to ensure data convergence.\\

\par IT functions of Ellis and Sun do not satisfy the property TP1 (see Fig.\ref{fig:vEllis} and Fig.\ref{fig:vSun}) \cite{imine02b}. The pairs of concurrent operations violating TP1 are $(o_1=Ins(1,f,pr_1),o_2=Del(1))$ and $(o_1=Ins(1,f),o_2=Del(1))$, respectively. \\

\begin{figure}[htbp]
\centering \vspace{-5mm}
\footnotesize 
\begin{minipage}[t]{0.9\linewidth}
\centerline{\xymatrix@C=10pt@M=2pt@R=10pt{ *+[F-,]\txt{site 1 \\
``efecte''} \ar@{.}'[d]'[dd]'[ddd][dddd] &
*+[F-,]\txt{site 2 \\ ``efecte''} \ar@{.}'[d]'[dd]'[ddd][dddd] \\
o_1=Ins(1,f,pr_1) \ar[ddr]  & o_2=Del(1) \ar[ddl] |!{[l];[dd]}\hole \\
*+[F]{\txt{``effecte''}} & *+[F]{\txt{``eecte''}} \\
  IT(o_2,o_1)=Del(2)     &  IT(o_1,o_2)= Ins(0,f,pr_1) \\
*+[F]{\txt{``efecte''}}  & *+[F]{\txt{``feecte''}} \\
}}
\caption{Violation of TP1 for Ellis{'}s IT.} 
\label{fig:vEllis}
\hfill
\end{minipage}

 \begin{minipage}[t]{0.9\linewidth}
\centerline{\xymatrix@C=10pt@M=2pt@R=10pt{ *+[F-,]\txt{site 1 \\
``efct''} \ar@{.}'[d]'[dd]'[ddd][dddd] &
*+[F-,]\txt{site 2 \\ ``efct''} \ar@{.}'[d]'[dd]'[ddd][dddd] \\
o_1=Ins(1,f) \ar[ddr]  & o_2=Ins(1,e) \ar[ddl] |!{[l];[dd]}\hole \\
*+[F]{\txt{``effct''}} & *+[F]{\txt{``efect''}} \\
IT(o_2,o_1)=Ins(2,e)     & IT(o_1,o_2)=Ins(2,f) \\
*+[F]{\txt{``efefct''}}  & *+[F]{\txt{``effect''}} \\} }
  \caption{Violation of TP1 for Sun{'}s IT.}
  \label{fig:vSun}
 \end{minipage} \normalsize \vspace{-5mm}
\end{figure}

\begin{figure}[htbp]
\centering
\footnotesize 
\begin{minipage}[t]{1\linewidth} 
\centerline{\xymatrix@C=10pt@M=2pt@R=10pt{ 
*+[F-,]\txt{site of $u_1$ \\ ``eftte''} \ar@{.}'[d]'[dd]'[ddd][dddd] &
*+[F-,]\txt{site of $u_2$ \\ ``eftte''} \ar@{.}'[d]'[dd]'[ddd][dddd] &
*+[F-,]\txt{site of $u_3$ \\ ``eftte''} \ar@{.}'[d]'[dd]'[ddd][dddd] &
*+[F-,]\txt{site of $u_4$ \\  ``eftte''} \ar@{.}'[d]'[dd]'[ddd][dddd] \\
o_1=Ins(3,f,\emptyset,\emptyset) \ar[ddddrr]  & o_2=Ins(2,c,\emptyset,\emptyset)  \ar[dddddddr]  &  o_{3}=Del(2)     & o_{3}=Del(2) \\
& &  o_{4}=Ins(2,e,\emptyset,\emptyset)     & o_{4}=Ins(2,e,\emptyset,\emptyset) \\
& & o_{5}=Del(2)     & o_{5}=Del(2) \\
& & *+[F]{\txt{``efte''}} \ar@{.}'[d]'[dd]'[ddd][dddd] & *+[F]{\txt{``efte''}}\ar@{.}'[d]'[dd]'[ddd][dddd] \\
& & o_1'=IT^*(o_1,[o_{3};o_{4};o_5]) & o_2'=IT^*(o_2,[o_{3};o_{4};o_5])\hole\\
& & o_1'=Ins(2,f,\{o_{3}\},\{o_{5}\}) \ar[ddr]  &  o_2'=Ins(2,c, \{o_{5}\}, \{o_{3}\}) \ar[ddl] |!{[l];[dd]}\hole \\
& & *+[F]{\txt{``effte''}} & *+[F]{\txt{``efcte''}} \\
& & o_2'=IT^*(o_2,[o_{3};o_{4};o_5]) & o_1'=IT^*(o_1,[o_{3};o_{4};o_5])\hole\\
& &  IT(o_2',o_1')= Ins(3,c,\{o_{5}\}, \{o_{3}\})     &   IT(o_1',o_2')=Ins(3,f,\{o_{3}\}, \{o_{5}\}) \\
& & *+[F]{\txt{``effcte''}}  & *+[F]{\txt{``efcfte''}} \\
}} \vspace{-2mm}
\caption{Violation of TP1 for Suleiman{'}s IT.}
\label{fig:vSuleiman} \vspace{-5mm}
\end{minipage} \normalsize
\end{figure}
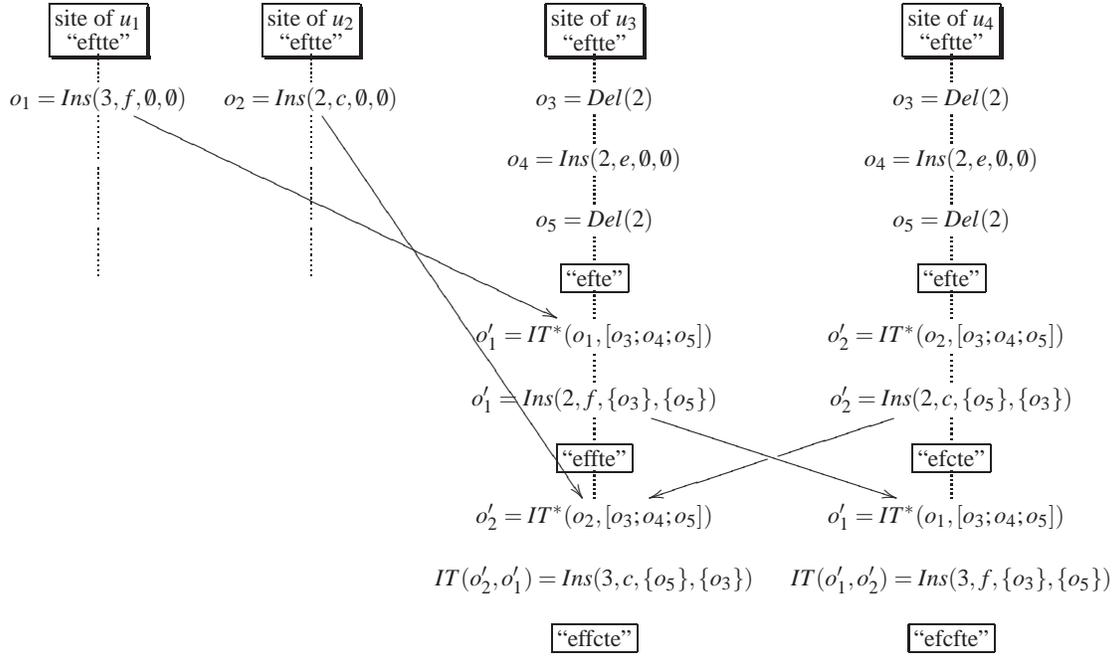

\par Suleiman's IT satisfies neither TP1 nor TP2 \cite{imine02b,IFM10}. The counterexample for TP1 is given by the pair of operations $(o_1'=Ins(2,f,\{o_3\},\{o_5\}),o_2'=Ins(2,c,\{o_5\},\{o_3\}))$. The corresponding scenario, reported at Fig.\ref{fig:vSuleiman}, consists of $4$ users $u_1, u_2, u_3$ and $u_4$ on different sites. Users $u_1$, $u_2$ and $u_3$ have generated and executed locally sequences  $S_1=[o_{1}=Ins(3,f,\emptyset,\emptyset)]$, $S_2=[o_{2}=Ins(2,c,\emptyset,\emptyset)]$ and $S_3=[o_{3}=Del(2); o_{4}=Ins(2,e,\emptyset,\emptyset); o_{5} = Del(2)]$, respectively. Then, user $u_3$ receives successively operations $o_1$ and $o_2$. User $u_4$ receives consecutively operations of $S_3$, $o_2$ and $o_1$. The IT function of Suleiman fails to ensure convergence (property TP1 is violated). Indeed, when the site of user $u_3$ receives $o_1$, it is first transformed w.r.t. the sequence $S_3$. The resulting operation $o_1'= IT^*(o_1, S_3)= Ins(3,f,\{o_{3}\},\{o_{5}\})$ is executed locally. When it receives $o_2$, it is successively transformed w.r.t. $S_3$ ($o_2'= IT^*(o_2, S_3)= Ins(2,c,\{o_{5}\},\{o_{3}\})$) and  $o_1'$ (i.e., $IT(o_2',o_1')=Ins(3,f,\{o_{3}\},\{o_{5}\})$) before its execution. For its part, the site of $u_4$ executes the sequence $S_3$ of $u_3$ without transformation but when it receives $o_2$, it is transformed against $S_3$ (i.e.,$o_2'= IT^*(o_2, S_3)= Ins(2,c,\{o_{5}\},\{o_{3}\})$) then executed. When it receives operation $o_1$, it is successively transformed w.r.t. $S_3$ (i.e., $o_1'$) and $o_2'$  (i.e., $IT(o_1',o_2')$) before its execution. This scenario leads to a divergence of copies of $u_3$ and $u_4$. The property $TP1$ is then violated.  \\

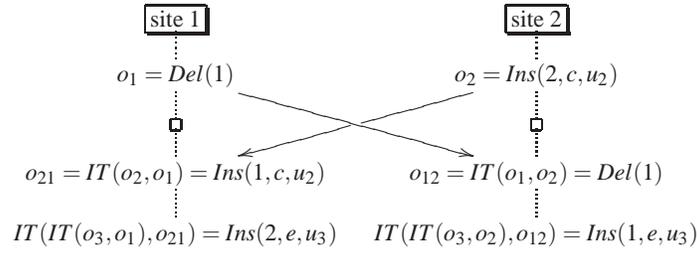
\begin{figure}
\centering
\footnotesize 
\begin{minipage}[t]{0.9\linewidth}
\centerline{\xymatrix@C=10pt@M=2pt@R=10pt{ *+[F-,]\txt{site 1} \ar@{.}'[d]'[dd]'[ddd][dddd] & *+[F-,]\txt{site 2} \ar@{.}'[d]'[dd]'[ddd][dddd] \\
o_1=Del(1) \ar[ddr]  & o_2=Ins(2,c,u_2) \ar[ddl] |!{[l];[dd]}\hole \\
*+[F]{} & *+[F]{} \\
o_{21}=IT(o_2,o_1)=Ins(1,c,u_2)     &   o_{12}=IT(o_1,o_2)=Del(1) \\
IT( IT(o_3,o_1), o_{21})=Ins(2,e,u_3)     &   IT( IT(o_3,o_2), o_{12})=Ins(1,e,u_3) \\
}} 
\caption{Violation of TP2 for Ressel{'}s IT (in case $u_2 < u_3$).} 
\label{fig:vRessel}
\end{minipage} \normalsize
\end{figure}

\par Ressel's IT does not satisfy TP2 but satisfies TP1 \cite{IFM10}. In Fig.\ref{fig:vRessel}, we report a scenario violating property TP2 for the triplet of concurrent operations $(o_1=Del(1), o_2=Ins(2,c_2,u_2), o_3=Ins(1,c_3,u_3))$. \\

\par
\begin{figure}
\footnotesize 
 \begin{minipage}[t]{0.9\linewidth} 
\centerline{\xymatrix@C=20pt@M=2pt@R=10pt{ 
*+[F-,]\txt{site of $u_1$ \\ ``eefft''} \ar@{.}'[d]'[dd]'[ddd][dddd] &
*+[F-,]\txt{site of $u_2$ \\ ``eefft''} \ar@{.}'[d]'[dd]'[ddd][dddd] &
*+[F-,]\txt{site of $u_4$ \\ ``eefft''} \ar@{.}'[d]'[dd]'[ddd][dddd] &
*+[F-,]\txt{site of $u_3$ \\ ``eefft''} \ar@{.}'[d]'[dd]'[ddd][dddd] \\
o_1=Del(2) \ar[ddr] & o_{0}=Del(1)     & o_{0}=Del(1) & o_3=Ins(2,e,2) \ar[ddddddl]\\
& *+[F]{\txt{``efft''}}\ar@{.}'[d]'[dd]'[ddd][dddd]  & *+[F]{\txt{``efft''}} \ar@{.}'[d]'[dd]'[ddd][dddd] & \\
& o_1'=IT(o_1,o_{0})=Del(1) \ar[ddr]  & o_{2}=Ins(2,c,2) \ar[ddl] |!{[l];[dd]}\hole  & \\
& *+[F]{\txt{``eft''}} \ar@{.}'[d]'[dd]'[ddd][dddd]&    *+[F]{\txt{``efcft''}}\ar@{.}'[d]'[dd]'[ddd][dddd] & \\
& o_{2}'=IT(o_{2},o_1')=Ins(1,c,2)     & o_1''= IT^*(o_1,[o_{0};o_{2}])=Del(1) & \\
& *+[F]{\txt{``ecft''}}  & *+[F]{\txt{``ecft''}}  & \\
& IT^*(o_3,[o_{0};o_1';o_{2}'])=Ins(2,e,2)     & IT^*(o_3,[o_{0};o_{2};o_1''])=Ins(1,e,2) & \\
& *+[F]{\txt{``eceft''}}  & *+[F]{\txt{``eecft''}}  & \\ }}
  \caption{Violation of TP2 for Imine{'}s IT.}
  \label{fig:vImine} \vspace{-6mm}
 \end{minipage} \normalsize
\end{figure}
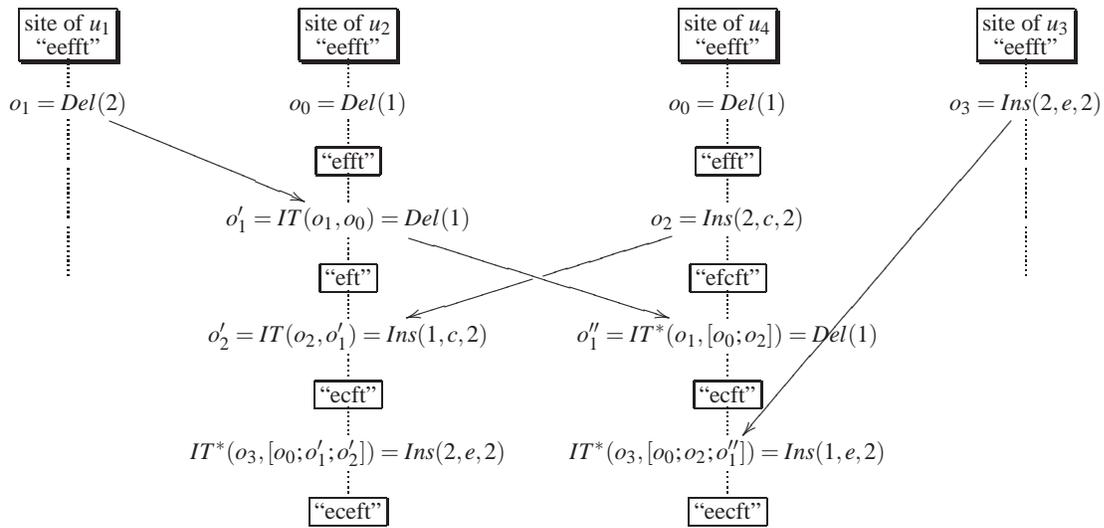

 Imine's IT function satisfies TP1 but does not satisfy TP2 \cite{IFM10}. In Fig.\ref{fig:vImine}, we report a scenario violating TP2. In this scenario, there are $4$ users $u_1, u_2, u_3$ and $u_4$ on different sites. Users $u_1$, $u_2$ and $u_3$ have generated sequences  $S_1=[o_{1}=Del(2)]$, $S_2=[o_{0}=Del(2);o_{2}=Ins(2,c,2)]$ and $S_3=[o_{3}=Ins(2,e,2)]$, respectively. User $u_2$ executes operations $o_{0}$ and $o_{2}$ then it receives successively operations $o_1$ and $o_3$. User $u_4$ receives successively operations $o_{0}$, $o_1$, $o_{2}$ and $o_3$. For this scenario, the IT function of Imine fails to ensure convergence for copies of users $u_2$ and $u_4$. The property $TP2$ is violated (see Fig.\ref{fig:vImine}).  

 
\begin{figure*}
\centering
\footnotesize
 \begin{minipage}[t]{0.45\linewidth} 
\includegraphics[width=1\textwidth]{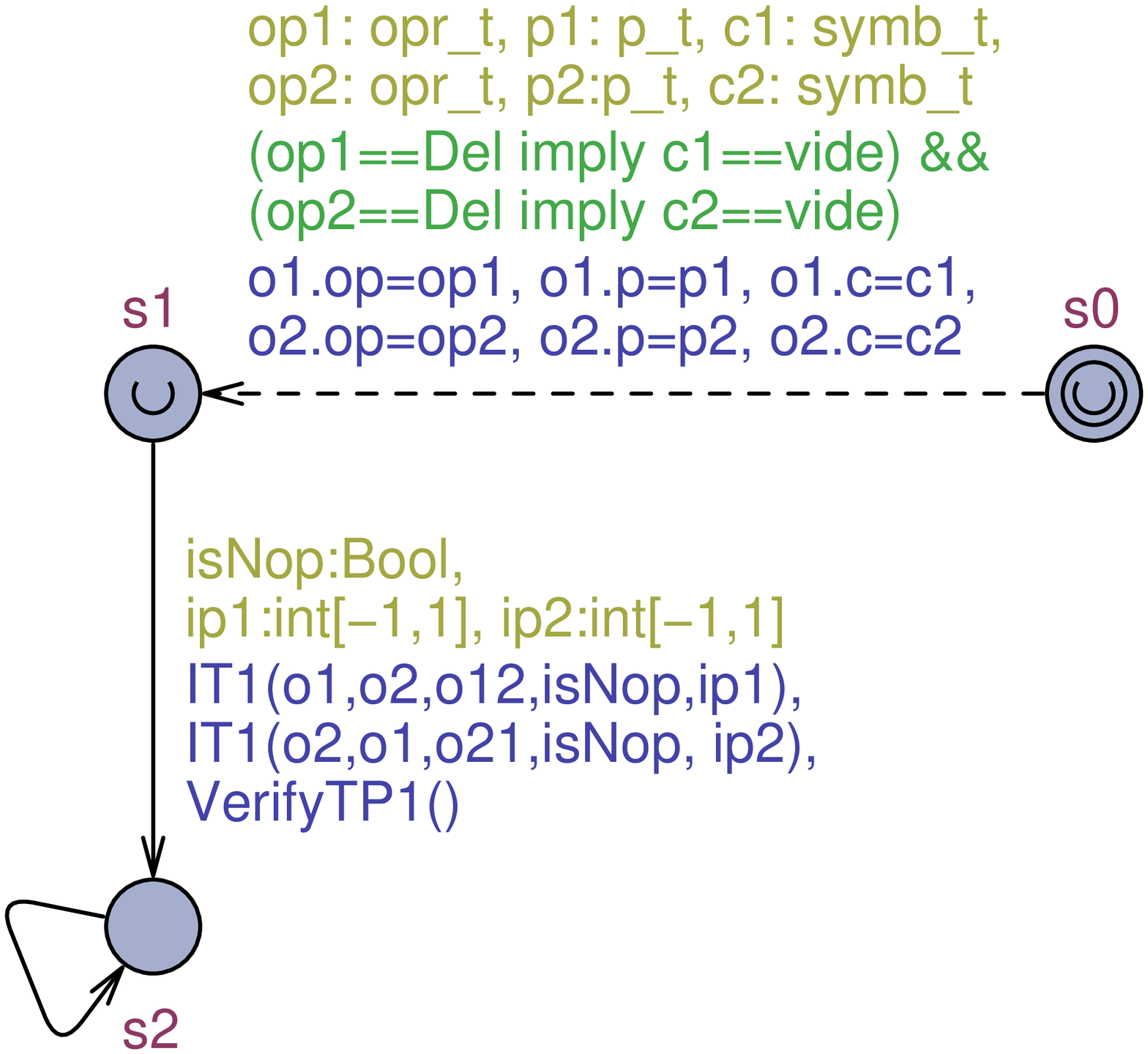} \vspace{-5mm}
 \caption{Synthesize an IT for TP1} \label{GATP1}
\end{minipage}
 \begin{minipage}[t]{0.45\linewidth}
\includegraphics[width=1\textwidth]{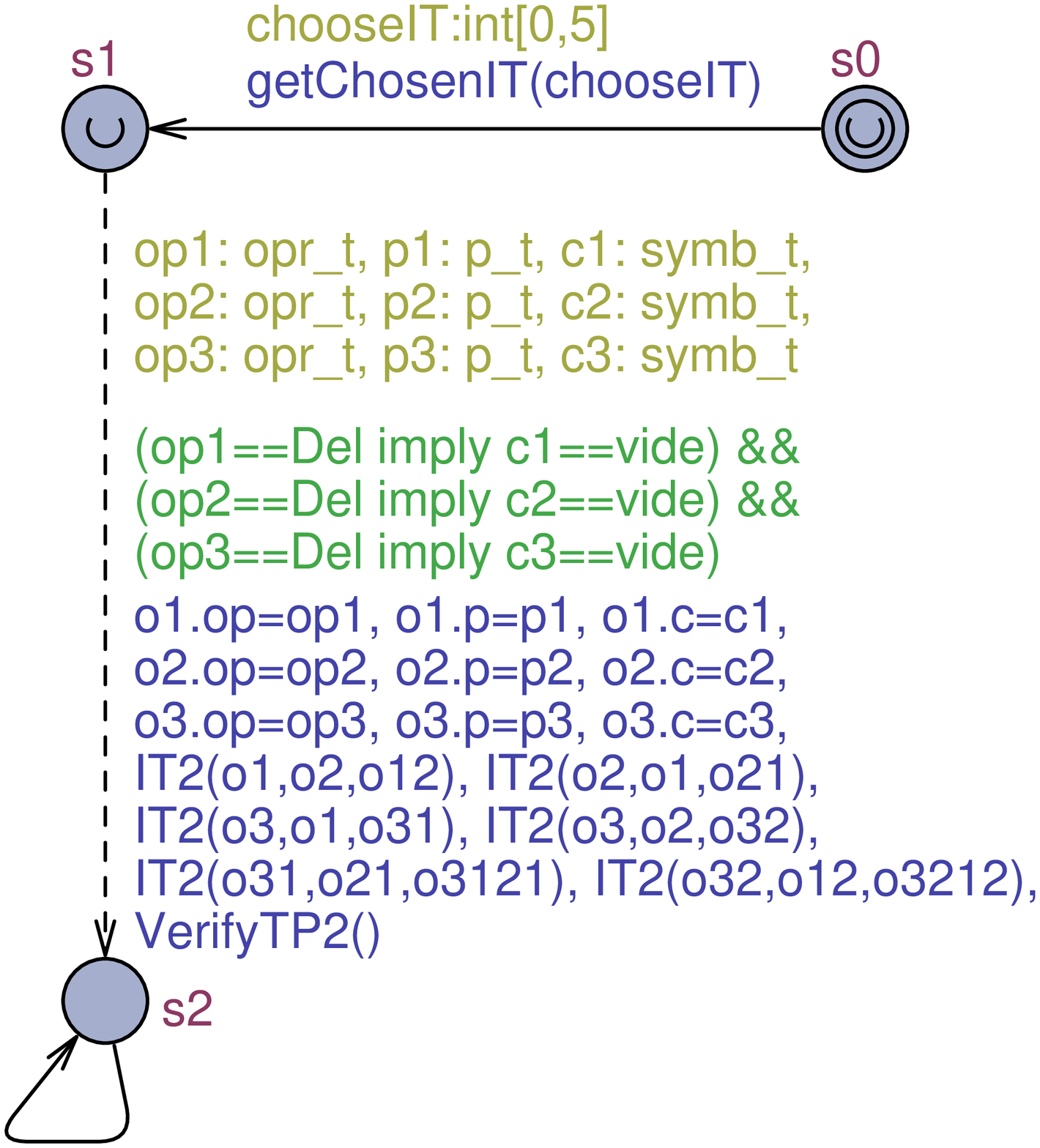} 
  \caption{Synthesize a consistent IT function} \label{GATP2} 
\end{minipage}\normalsize
\end{figure*}

\vspace{-5mm}
\section{Controller synthesis of consistent IT functions} \label{s3} \vspace{-3mm}
Given the model of some system and a property to be satisfied. Controller synthesis addresses the question of how to limit the  behavior of the model so as to meet the property. In such a framework, the model consists, in general, of controllable and uncontrollable actions (i.e., transitions). The control objective is to find, if it exists, a strategy to force the property, by choosing appropriately controllable actions to be executed, no matter what uncontrollable actions are executed. We are interested to apply the principle of controller synthesis to design an IT function which satisfies properties TP1 and TP2. We first investigate whether or not there exist some IT functions which satisfy property TP1. If it is the case, we investigate  whether or not there exist some IT functions, among those satisfying TP1, which satisfy also TP2. \\
    
\par For these investigations, we use the game automata formalism `\`a la UPPAAL' \cite{Larsen}. A game automaton is an automaton with two kinds of transitions: controllable and uncontrollable. Each transition has a source location and a destination location. It is annotated with selections, guards and blocks of actions. Selections bind non-deterministically a given identifier to every value in a given range (type). The other labels of a transition are within the scope of this binding. A state is defined by the current location and the current values of all variables. A transition is enabled in a state iff the current location is the source location of the transition and its guard evaluates to true.  The firing of the transition consists in reaching its destination location and executing atomically its block of actions. The side effect of this block changes the state of the system. To force some properties, the enabled transitions that are controllable can be delayed or simply ignored.   However, the uncontrollable transitions can neither be delayed nor ignored.

\subsection{Do there exist IT functions which satisfy TP1?} 
An IT function satisfies property $TP1$ iff for any pair of concurrent operations $o_1$ and $o_2$, it holds that $[o_1; IT(o_2;o_1)] \equiv [o_2; IT(o_1,o_2)]$. To verify whether or not there are some IT functions which satisfy property TP1, we have represented in the game automaton, depicted at Fig.\ref{GATP1}, the generation of operations $o_1$ and $o_2$, the computation of $IT(o_1;o_2)$ and $IT(o_2,o_1)$,  and the verification of $[o_1; IT(o_2;o_1)] \equiv [o_2; IT(o_1,o_2)]$.
The generation of operations is specified by the uncontrollable transition $(s_0,s_1)$, since we have no control on the kinds operations generated by users. The operational transformations and the verification of TP1 are represented by the controllable transition $(s_1,s_2)$. 
The model starts by selecting two operations $o_1$ and $o_2$. The domain of operations is fixed so as to cover all cases of transformations. 
Afterwards, it chooses two transformations to apply to $o_1$ w.r.t. $o_2$ and $o_2$ w.r.t. $o_1$ and applies them by invoking function $IT1$. Function $IT1(o_1,o_2,o_{12},IsNop,ip_1)$ returns in $o_{12}$ the result of transformation of $o_1$ w.r.t. $o_2$. If $IsNop=false$ then $o_{12}=Nop()$, otherwise the transformation of $o_1$ consists in updating the parameter position ($o_{12}.p=o_1.p+ip_1$). It means that $4$ possibilities are offered for transforming an operation $o_1$ w.r.t. another  operation $o_2$: Nop(), decrementing, maintaining, or incrementing the position of $o_1$. 
Finally, the model verifies whether or not the property TP1 is satisfied. No matter what operations $o_1$ and $o_2$ generated by the uncontrollable transition, the controller synthesis aims to force property TP1 by choosing appropriately the operational transformations. \\
\par We have used the tool \emph{Uppaal-Tiga} \cite{Larsen} to verify whether or not there exist some IT functions, which satisfy TP1. The safety control objective for TP1 is $AG \ TP1$, where $TP1$ is defined in the model as a boolean variable whose value is  $true$ while the property TP1 is satisfied. The boolean variable TP1 is set to false by the function VerifyTP1 if  $[o_1;IT(o_2,o_1)] \not \equiv [o_2;IT(o_1,o_2)]$. \emph{Uppaal-Tiga} concludes that the property is satisfied, which means that there is, at least, a strategy to force property TP1. We report in Table \ref{tab:TP1} the different IT functions (satisfying TP1) extracted from the output file of the tool \emph{verifytga} of \emph{Uppaal-Tiga}.\\

\par Even if some operational transformations satisfy TP1, they are unacceptable from the semantic point of view. For instance, if $p_1=p_2$, the operational transformations $IT(Del(p_1), Del(p_2))= Del(p_1-1)$, $IT(Del(p_1), Del(p_2))= Del(p_1)$ and $IT(Del(p_1), Del(p_2))= Del(p_1+1)$ mean that if two users generate concurrently the same delete operation, two symbols will be deleted in each site, which is unacceptable from the semantic point of view. The only operational transformation which has a sense for this case is  $IT(Del(p_1), Del(p_2))=Nop()$. It means that only the symbol at position $p_1$ is deleted in each site. After eliminating these incoherent operational transformations, it remains $2$ possibilities for $IT(Ins(p_1,c_1),Ins(p_2,c2)), p_1=p_2, c_1 \neq c_2$, and $3$ for $IT(Ins(p_1,c_1),Ins(p_2,c2)), p_1=p_2, c_1 = c_2$. Therefore, we can extract $6$ IT functions which satisfy TP1. These IT functions differ in the way that conflicting operations are managed. 
\begin{table*}[htbp]
\caption{IT functions supplied by \emph{Uppaal-Tiga} for TP1 and classical signatures of update operations}
\label{tab:TP1}
\centering 
\footnotesize 
\begin{tabular}{|c|c||c||c|c|}
\hline 
$o_1$ & $o_2$ & $Cnd(p_1,p_2,c_1,c_2)$ & $IT(o_1,o_2)$& $IT(o_2,o_1)$\\\hline \hline
$Ins(p_1,c_1)$ & $Ins(p_2,c_2)$ & $p_1 < p_2$ & $Ins(p_1,c_1)$& $Ins(p_2+1,c_2)$\\\hline 

\textcolor{red}{ $Ins(p_1,c_1)$} & \textcolor{red}{$Ins(p_2,c_2)$} & \textcolor{red}{$p_1=p_2 \wedge c_1 < c_2$} & \textcolor{red}{$Ins(p_1+1,c_1)$}& \textcolor{red}{$Ins(p_2,c_2)$}\\\hline
\textcolor{red}{$Ins(p_1,c_1)$} & \textcolor{red}{$Ins(p_2,c_2)$} & \textcolor{red}{$p_1=p_2 \wedge c_1 < c_2$} & \textcolor{red}{$Ins(p_1,c_1)$}& \textcolor{red}{$Ins(p_2+1,c_2)$}\\\hline

\textcolor{blue}{$Ins(p_1,c_1)$} & \textcolor{blue}{$Ins(p_2,c_2)$} & \textcolor{blue}{$p_1=p_2 \wedge c_1=c_2$} & \textcolor{blue}{$Ins(p_1+1,c_1)$}& \textcolor{blue}{$Ins(p_2+1,c_2)$}\\\hline
\textcolor{blue}{$Ins(p_1,c_1)$} & \textcolor{blue}{$Ins(p_2,c_2)$} & \textcolor{blue}{$p_1=p_2 \wedge c_1=c_2$} & \textcolor{blue}{$Ins(p_1,c_1)$}& \textcolor{blue}{$Ins(p_2,c_2)$}\\\hline
\textcolor{blue}{$Ins(p_1,c_1)$} & \textcolor{blue}{$Ins(p_2,c_2)$} & \textcolor{blue}{$p_1=p_2 \wedge c_1=c_2$} & \textcolor{blue}{$Nop()$}& \textcolor{blue}{$Nop()$}\\\hline\hline

$Del(p_1)$ & $Del(p_2)$ & $p_1 < p_2$& $Del(p_1)$& $Del(p_2-1)$\\\hline 
\textcolor{green}{$Del(p_1)$} &\textcolor{green}{$Del(p_2)$} & \textcolor{green}{$p_1=p_2$} & \textcolor{green}{$Del(p_1-1)$}& \textcolor{green}{$Del(p_2-1)$}\\\hline 
\textcolor{green}{$Del(p_1)$} &\textcolor{green}{$Del(p_2)$} & \textcolor{green}{$p_1=p_2$} & \textcolor{green}{$Del(p_1+1)$}& \textcolor{green}{$Del(p_2+1)$}\\\hline 
\textcolor{green}{$Del(p_1)$} & \textcolor{green}{$Del(p_2)$} & \textcolor{green}{$p_1=p_2$} & \textcolor{green}{$Del(p_1)$}& \textcolor{green}{$Del(p_2)$}\\\hline 
\textcolor{green}{$Del(p_1)$} & \textcolor{green}{$Del(p_2)$} & \textcolor{green}{$p_1=p_2$} & \textcolor{green}{$Nop()$}& \textcolor{green}{$Nop( )$}\\\hline \hline

$Ins(p_1,c_1)$ & $Del(p_2)$ & $p_1 < p_2$& $Ins(p_1,c_1)$& $Del(p_2+1)$\\\hline 
$Ins(p_1,c_1)$ & $Del(p_2)$ & $p_1=p_2$& $Ins(p_1,c_1)$& $Del(p_2+1)$\\\hline \hline

$Del(p_1)$ & $Ins(p_2,c_2)$ & $p_1 < p_2$& $Del(p_1)$& $Ins(p_2-1,c_2)$\\\hline 
$Del(p_1)$ & $Ins(p_2,c_2)$ & $p_1=p_2$ & $Ins(p_1,c_1)$& $Del(p_2+1)$\\\hline 
\end{tabular}
\end{table*} 

\subsection{Do there exist IT functions which satisfy TP1 and TP2?} 
An IT function satisfies property $TP2$ iff for any triplet of pairwise concurrent operations $o_1$, $o_2$ and $o_3$, it holds that $IT( IT(o_3,o_1), IT(o_2,o_1) ) =  IT( IT(o_3,o_2), IT(o_1,o_2))$. To verify whether or not there are some IT functions which satisfy properties TP1 and TP2, we have used the game automaton depicted at Fig.\ref{GATP2}.  This model starts by selecting an IT function, which satisfies property TP1 (the range of $chooseIT$ corresponds to the $6$ IT functions satisfying TP1). Afterwards, it selects three operations $o_1$, $o_2$ and $o_3$, and performs the transformations needed to verify TP2. Function $IT2(o_1,o_2,o_{12})$ applies the selected IT function to $o_1$ w.r.t. $o_2$ and returns the result of this transformation in $o_{12}$.  Finally, the model calls function  VerifyTP2. The control aims to force to choose the appropriate IT function so as to satisfy property TP2. The control objective is specified by the CTL formula $AG \ TP2$, where $TP2$ is  a boolean variable whose value is $true$ while the property TP2 is satisfied. This variable is set to false by the function VerifyTP2 if $IT(IT(o_3,o_1),IT(o_2,o_1)) \neq IT(IT(o_3,o_2),IT(o_1,o_2)).$\\

\par \emph{Uppaal-Tiga} concludes that the property $AG \ TP2$ cannot be forced, which means that  there is no strategy to force property TP2. In other words, there is no IT function, based on classical parameters of delete and insert operations, which satisfies both TP1 and TP2. We have investigated why there is no consistent IT function based on the basic parameters of delete and insert operations. This investigation has led to isolate two symbolic pairwise scenarios which prevent from getting a consistent IT function. We report in Fig.\ref{fig:Scen1} and Fig.\ref{fig:Scen2} these two pairwise sequences named scenario 1 and scenario 2, respectively. 
For scenario 1, to verify TP2, the computed operational transformations are: \\
$o_{21}=IT(o_2,o_1)= IT(Ins(p_1,c_2) ,o_1) = Ins(p_1,c_2)$,\\
 $o_{12}=IT(o_1,o_2)= IT(Del(p_1),Ins(p_1,c_2))=Del(p_1+1)$,\\
 $o_{31}=IT(o_3,o_1)=Ins(p_1,c_3)$,  \ \ \ \ 
 $o_{32}=IT(o_3,o_2)=Ins(p_1+2,c_3)$, \\ 
 $IT(o_{32},o_{12})=IT(Ins(p_1+2,c_3),Del(p_1+1))= Ins(p_1+1,c_3)$ and \\
 $IT(o_{31},o_{21})=IT(Ins(p_1,c_3),Ins(p_1, c_2))$. \\
 For the last transformation, we have different possibilities (see Table \ref{tab:TP1}). To satisfy TP2, we must choose $IT(Ins(p_1,c_3),Ins(p_1, c_2))=Ins(p_1+1,c_3)$. \\ 
For scenario 2, the computed operational transformations are: \\
$o_{21}=IT(o_2,o_1)=Ins(p_1,c_2)$,  \ \ \ \ 
$o_{12}=IT(o_1,o_2)=Del(p_1)$, \\ 
$o_{31}=IT(o_3,o_1)=Ins(p_1,c_3)$,  \ \ \ \ 
$o_{32}=IT(o_3,o_2)=Ins(p_1,c_3)$,\\ 
 $IT(o_{32},o_{12})=IT(Ins(p_1,c_3),Del(p_1))= Ins(p_1,c_3)$ and \\ 
 $IT(o_{31},o_{21})=IT(Ins(p_1,c_3),Ins(p_1, c_2))$.  \\
 To satisfy TP2, for the last operational transformation, we must use $IT(Ins(p_1,c_3),Ins(p_1, c_2))=Ins(p_1,c_3)$.\\
\par Consequently, a consistent IT function, if it exists, must have additional parameters in its operation signatures.  We have seen, in the previous section, different IT functions based on extending the insert signature with priority, issuer site, initial position or sets of deleted symbols before and after the position of the operation. We have reported divergent scenarios for all these IT functions. It means that the suggested additional parameters are not sufficient or appropriate to ensure convergence. Indeed, adding priority (as in Ellis's IT) or owner identifier (as in Ressel's IT) to the insert signature fails to ensure convergence for scenarios 1 and 2. Scenario 1 violates TP1 for Ellis's IT  (see Fig.\ref{fig:vEllis}). Scenario 2 violates TP2 for Ressel's IT (see Fig.\ref{fig:vRessel}).  For Suleiman's IT and Imine's IT, scenarios 1 and 2 satisfy TP1 and TP2 but the added parameters introduce other cases of divergence.

\begin{figure*}
\begin{center}
\footnotesize 
 \begin{minipage}[t]{0.9\linewidth}
\centerline{\xymatrix@C=10pt@M=2pt@R=10pt{ *+[F-,]\txt{site 1} \ar@{.}'[d]'[dd]'[ddd][dddd] &
*+[F-,]\txt{site 2} \ar@{.}'[d]'[dd]'[ddd][dddd] \\
o_1=Del(p_1) \ar[ddr]  & o_2=Ins(p_1,c_2) \ar[ddl] |!{[l];[dd]}\hole \\
*+[F]{} & *+[F]{} \\
 o_2=Ins(p_1,c_2)     &   o_1=Del(p_1) \\
o_3=Ins(p_1+1,c_3)     &   o_3=Ins(p_1+1,c_3) \\
}} 
  \caption{Scenario 1}
  \label{fig:Scen1} 
\hfill
 \end{minipage} 

 \begin{minipage}[t]{0.9\linewidth}
\centerline{\xymatrix@C=20pt@M=2pt@R=10pt{ *+[F-,]\txt{site 1} \ar@{.}'[d]'[dd]'[ddd][dddd] &
*+[F-,]\txt{site 2} \ar@{.}'[d]'[dd]'[ddd][dddd] \\
o_1=Del(p_1) \ar[ddr]  & o_2=Ins(p_1+1,c_2) \ar[ddl] |!{[l];[dd]}\hole \\
*+[F]{} & *+[F]{} \\
o_2=Ins(p_1+1,c_2)     & o_1=Del(p_1) \\
o_3=Ins(p_1,c_3)     &   o_3=Ins(p_1,c_3)}} 
  \caption{Scenario 2}
  \label{fig:Scen2} \vspace{-5mm}
\hfill
 \end{minipage} \normalsize
 \end{center}
\end{figure*}

\section{Conclusion} \label{s4}
In this work, we tried to answer the following question: what are all possible IT functions ensuring convergence for shared strings altered by insert and delete operations?
 We have first formulated the existence  problem of  a consistent IT function as a synthesis controller problem. As a main contribution, we have shown that only TP1 is satisfied
by some IT functions based on the position and character parameters. Thus, it is impossible to meet TP2 with these simple signatures. \\

\par Accordingly, the position and character parameters are necessary but not sufficient. In other words, additional parameters are needed to explore the existence of consistent IT
functions. In the near future,  we will follow the same framework to deal with the following issue: what are the minimal number of extra parameters to be added in order to achieve
consistent IT functions?

\bibliographystyle{eptcs}
\bibliography{mybib}
\end{document}